\documentclass[a4paper,11pt]{article}
\pdfoutput=1 

\usepackage{jcappub} 

\usepackage[T1]{fontenc} 
\usepackage{enumitem}

\def\be{\begin{equation}}
\def\ee{\end{equation}}
\def\ben{\begin{eqnarray}}
\def\een{\end{eqnarray}}
\def\bes{\begin{subequations}}
\def\ees{\end{subequations}}

\def\bk{{\bf k}}

\def\br{{\bf r}}

\def\bk{{\bf k}}

\def\bx{{\bf x}}
\def\2p{{(2\pi)^2}}

\def\be{\begin{equation}}
\def\ee{\end{equation}}
\def\beq{\begin{equation}}
\def\eeq{\end{equation}}
\def\ben{\begin{eqnarray}}
\def\een{\end{eqnarray}}

\newcommand{\beqa}{\begin{eqnarray}}
\newcommand{\eeqa}{\end{eqnarray}}

\def\ikap0{{\cal J}_{\theta_0}(r)}

\def\one1{\langle \kappa_{(i)}\kappa_{(j)} \rangle}
\def\one{{[\bar \xi^{(ij)}]}}

\def\ba{\begin{eqnarray}}
\def\ea{\end{eqnarray}}

\def\bk{{\bf k}}
\def\bq{{\bf q}}

\newcommand{\semibold}[1]{{\fontseries{b}\selectfont{#1}}}
\newcommand{\para}[1]{\par\vspace{2mm}\noindent\semibold{{#1.}---}\ignorespaces}

\renewcommand{\leq}{\leqslant}

\newcommand{\llangle}{\langle\kern-2\nulldelimiterspace\langle}
\newcommand{\rrangle}{\rangle\kern-2\nulldelimiterspace \rangle}
\newcommand{\bn}{\mathbf{n}}
\newcommand{\bv}{\mathbf{v}}

\title{An Inventory of Bispectrum Estimators for Redshift Space Distortions}
\author{Donough Regan}
\affiliation{Astronomy Centre, School of Mathematical and Physical Sciences,\\ University of Sussex, Brighton BN1 9QH, U.K.}
\emailAdd{D.Regan@sussex.ac.uk}

\abstract{ In order to best improve constraints on cosmological parameters and on models of modified gravity using current and future galaxy surveys it is necessary maximally exploit the available data. As redshift-space distortions mean    statistical translation invariance is broken for galaxy observations, this will require measurement of the monopole, quadrupole and hexadecapole of not just the galaxy power spectrum, but also the galaxy bispectrum. A recent (2015) paper by Scoccimarro demonstrated how the standard bispectrum estimator may be expressed in terms of Fast Fourier Transforms (FFTs) to afford an extremely efficient algorithm, allowing the bispectrum multipoles on all scales and triangle shapes to be measured in comparable time to those of the power spectrum. In this paper we present a suite of alternative proxies to measure the three-point correlation multipoles. In particular, we describe a modal (or plane wave) decomposition to capture the information in each multipole in a series of basis coefficients, and also describe three compressed estimators formed using the skew-spectrum, the line correlation function and the integrated bispectrum, respectively. As well as each of the estimators offering a different measurement channel, and thereby a robustness check, it is expected that some (especially the modal estimator) will offer a vast data compression, and so a much reduced covariance matrix. This compression may be vital to reduce the computational load involved in extracting the available  three-point information.

\par\vspace{40mm}\mbox{}\hfill
\raisebox{-0.5\height}{\includegraphics[scale=0.2]{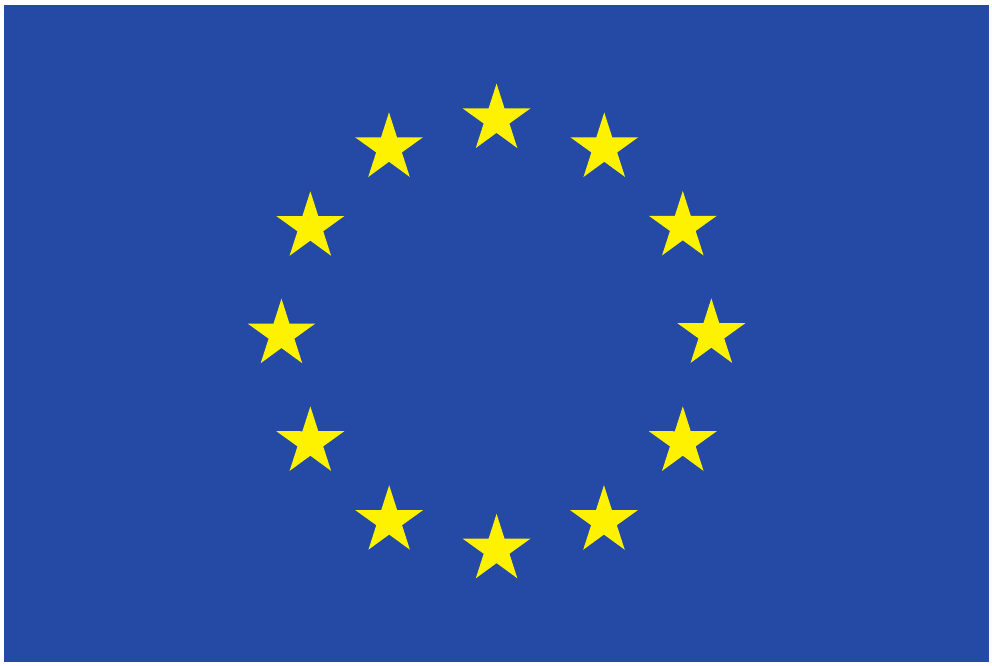}}
\;{\vrule width 1pt}\;
\raisebox{-0.5\height}{\includegraphics[scale=0.1]{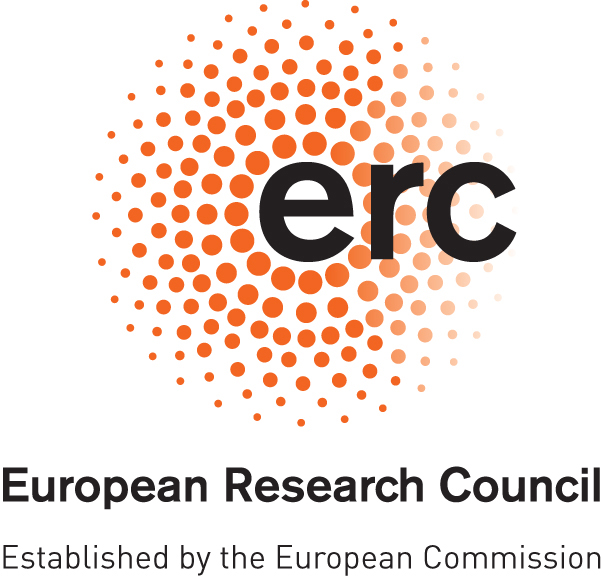}}
\quad
\raisebox{-0.5\height}{\includegraphics[scale=0.5]{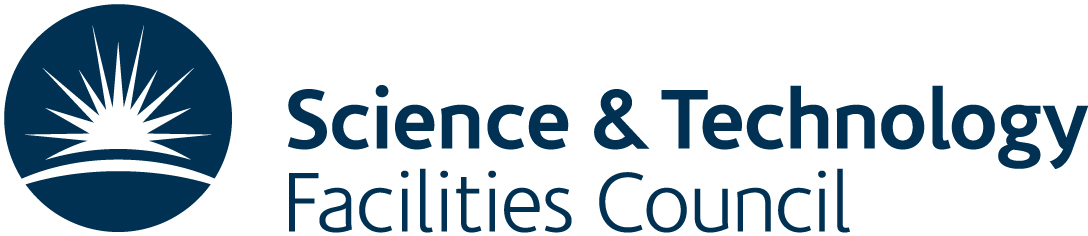}}}


\begin{document}
\maketitle
\flushbottom

\section{Introduction}
While the cosmic microwave background (CMB) has been the primary source of information on the early universe to date (c.f. \citep{Ade:2013sjv,Hinshaw:2012aka}), future advancements will undoubtedly be driven by large galaxy redshift surveys. The next decade promises huge datasets from the Euclid Spectroscopic instrument \citep{Laureijs:2011gra}, the Square Kilometre Array (SKA) \citep{Maartens:2015mra} and the Dark Energy Spectroscopic Instrument (DESI) \citep{Aghamousa:2016zmz}, but analysing this data presents challenges the CMB did not. 

From the {\it prediction} perspective, the non-linear evolution of matter, allied to the complications due to the biasing of galaxies from their dark matter hosts, mean that sophisticated modelling techniques must be developed. In addition, the vast increase in the data source means that historically popular algorithms to {\it estimate} the two point correlation function (or its Fourier equivalent, the power spectrum) using galaxy pair counts are unfeasible. In addition, as a non-linear source, one should consider statistical measures beyond the two-point function. Efficient and robust measurement of the three-point function from galaxy surveys is the key focus of this paper.

The measurement of the galaxy multispectra is complicated by the presence of peculiar velocities, $\bv$, owing to which observations of galaxies are displaced (with respect to the Hubble flow) from their actual positions, $\br$, relative to the observer to 
\begin{align}\label{eq:r2s}
\mathbf{s}=\br + \frac{\bv . \hat{\br}}{H} \hat{\br}\,.
\end{align}
Conservation of mass allows one to relate the redshift space matter overdensity, $\delta_s$, to the real space, $\delta$, \citep{Scoccimarro:2004tg}
\begin{align}\label{eq:Delta2Deltas}
\delta_s(\bk)&=\delta(\bk)+\int d^3 r e^{- i \bk.\hat{\br}}\left[\exp\left(-\frac{i}{H} (\bk. \hat{\br})(\bv(\br).\hat{\br})\right)\right](1+\delta(\br)) \\
&\approx \delta(\bk) -\frac{i}{H}(\bk.\hat{\br}) \hat{\br}.\bv(\bk)  -\frac{i}{H}(\bk.\hat{\br}) [\delta (\hat{\br}.\bv)]_{\bk} - \frac{(\bk.\hat{\br})^2}{2 H^2}\left[(1+\delta) (\hat{\br}.\bv)^2\right]_{\bk} + \dots  \nonumber
\end{align}
where we have employed the notation $[f]_\bk \equiv \int d^3 x e^{-i \bk. \bx} f(\bx)$. The first two terms constitute the usual Kaiser formula, $\delta_s(\bk)\approx \delta(\bk)(1+ f (\hat{\bk}.\hat{\br})^2)$, where the leading order peculiar velocity is given by $\bv(\bk) \approx i H f \delta(\bk) {\bk}/k^2$. As a probe of the peculiar velocity, redshift space distortions are especially interesting for testing models of modified gravity. Therefore, redshift space distortions, far from being a nuisance, offer an extremely valuable source of information. 

\para{Predicting the power spectrum and bispectrum} Analytic estimates for the two point correlation function (or its Fourier transform, the power spectrum) typically involve the use of perturbation theory, to express the redshift space power spectrum in terms of the real space power spectrum. In addition one uses the plane parallel approximation to replace $\hat{\br}$ in Eq.~\eqref{eq:r2s} by the line of sight $\hat{\bn}$ to the observer. While the plane parallel approximation is an accurate approximation for existing datasets, this approximation will be insufficient for future wide-angle surveys. A more pressing issue however is that if one uses Eq.~\eqref{eq:Delta2Deltas} to relate the real space power spectrum to the redshift space power spectrum within standard perturbation theory (SPT), the results diverge even on relatively large scales. This lack of agreement originates due the effect of large bulk flows which are not treated adequately in the standard (Eulerian) perturbation theory. Expressing the SPT result in the Lagrangian perturbation theory effectively resums these bulk velocity terms, thereby damping the impact of acoustic oscillations \citep{Vlah:2015sea,Vlah:2015zda,Senatore:2014via,delaBella:2017qjy}. Unfortunately the origin of this damping has often been conflated in the literature with the {\it Fingers-of-God} effect which arises due to random peculiar velocities of galaxies within virialised objects. 

Recent analytic frameworks such as the Effective Field Theory of Large Scale Structure \citep{Baumann:2010tm,Carrasco:2012cv} account for non-linear contributions by way of counterterms which aggregate the effect of UV physics, and whose amplitude are set by comparison to data/simulations. This framework has been applied to RSDs in \citep{Senatore:2014vja,Lewandowski:2015ziq,Perko:2016puo,delaBella:2017qjy}. Predictive modelling for the redshift space distorted bispectrum is much less developed, but encouraging progress has been made in this direction recently \citep{Tellarini:2016sgp,Hashimoto:2017klo}.

\para{Measuring the power spectrum and bispectrum}
As radial distortions preserve isotropy, the angular dependency is best decomposed using spherical harmonic transforms. While progress has been made in this direction -- using, for example, the Fourier-Laguerre transform \citep{McEwen:2013jpa,Leistedt:2013vea}  --, we restrict our focus in this paper to the small-angle limit for which a Fourier analysis suffices and has the advantage of computational efficiency. Evaluation of the line-of-sight dependent moments allows one to probe the clustering as a function of angle. Treating the line of sight (LOS) for pairs of galaxies as a single LOS (to, say, the pair centre) provides a much more computationally efficient estimator, with effects of this small-angle approximation being small on the scales of interest. Recent work in \citep{Scoccimarro:2015bla,Bianchi:2015oia} demonstrated that the two- and three-point functions may be measured using simple and highly efficient Fast Fourier Transforms (FFTs) by employing this approximation. Despite the great improvement in efficiency, incorporating the multipoles of the RSD bispectrum into cosmological constraints will require estimation of the covariance matrix. Given the non-linear scales involved, this generally requires the use of mock catalogues. The high number of possible configurations of the bispectrum that may be measured in principle renders the number of catalogues required unfeasibly high, unless one coarse-grains well beyond the fundamental frequency. This motivates the search for alternative proxies for measurement and/or robustness checks of the standard estimator. 

In \citep{Byun:2017fkz} a suite of estimators were considered for this purpose in the context of real-space dark matter simulations. The {\it modal estimator} was determined to provide a data compression with as few as $\mathcal{O}(10)$ basis modes giving comparable constraining power to the standard estimator. The {\it line correlation function} and {\it integrated bispectrum} proxies were also considered, with the former providing competitive, though sub-optimal constraining power, and the latter showing little dependence on the underlying gravitational evolution (which may prove useful should one be interested in, say, primordial non-Gaussianity). The aim of this paper is to describe how each of these proxies, as well as the {\it skew-spectrum} may be generalised to measure the RSD bispectrum multipoles.

\para{Outline}
We will review the techniques of \citep{Scoccimarro:2015bla,Bianchi:2015oia} for the power spectrum in Section~\ref{sec:powerspec}, and standard bispectrum estimator in Section~\ref{subsec:StdEst}. The remainder of Section~\ref{sec:bispec} represents the novel results of this paper. Beginning in Section~\ref{subsec:SkewSpec}, we describe the skew spectrum technique whereby the 3-point function is reduced to the correlation between a 2-point and 1-point function of the density field, thus reducing the corresponding statistic to a function of a single wavenumber. In the following two subsections, Section~\ref{subsec:LCF} and Section~\ref{subsec:IB} we describe how the line correlation function and integrated bispectrum statistics can be straightforwardly extended to measure the RSD bispectrum multipoles. Finally in Section ~\ref{subsec:Modal} we begin by briefly recapitulating the modal basis decomposition technique for real space density fluctuations, which has been demonstrated to provide a vast data compression in the case of N-body simulations with negligible impact on constraining power. In each subsection we begin by describing the real space estimator, followed by the extension to redshift space. Finally in Section~\ref{sec:conclu} we present our concluding remarks. While this paper does not deal with the implementation of these estimators, or the complexities involved in actual measurement, such as the effects of a complex survey geometry, the bispectrum multipole proxies presented in this paper should greatly increase the scope for efficient and robust measurement of the three point function using upcoming galaxy surveys.

\section{Power Spectrum Multipole Estimator}\label{sec:powerspec}

The redshift space distortion power spectrum defined as
\begin{align}
\langle  \delta_s(\bk_1)\delta_s(\bk_2)\rangle = (2\pi)^3 \delta_D(\bk_1+\bk_2) P_s(k_1)\,,
\end{align}
can be estimated using
\begin{align}
\hat{P}_s(k) = \int \frac{d^2 \hat{k}}{4\pi} \frac{|\delta_s(\bk)|^2}{V} = \frac{1}{V}\int d^3 x_1 d^3 x_2 e^{i (\bx_1-\bx_2).\bk} \delta_s(\bx_1)\delta_s(\bx_2)\,,
\end{align}
where one averages over the wavevector directions to get an estimate that depends on the wavenumber only. Here and throughout we denote with a hat the estimator for the relevant statistic. Due to the discretisation of the wavenumbers, one usually averages the power spectrum over some range of values $k\pm \Delta k/2$. 

\para{Multipoles of the RSD Power Spectrum}
Computation of the power spectrum multipoles proceeds similarly, via the prescription,
\begin{align}
\hat{P}_{s}^{ (\ell)}(k) = \frac{2\ell+1}{V}\int d^3 x_1 d^3 x_2 e^{i (\bx_1-\bx_2).\bk} \delta_s(\bx_1)\delta_s(\bx_2)\mathcal{L}_\ell(\hat{\bk}.\hat{{\mathbf{n}}})\,,
\end{align}
where $\mathcal{L}_\ell$ represents the Legendre polynomial of degree $\ell$. The plane parallel approximation allows one to freely choose the line of sight direction to satisfy $\hat{\bf{n}}\parallel \hat{\bx}_1$; this greatly simplifies the computation, such that
\begin{align}\label{eq:PowSpecMultipole}
\hat{P}_{s}^{ (\ell)}(k) = \frac{1}{V}\int \frac{d^2 \hat{k}}{4\pi} \delta_{s}^{ (\ell)}(\bk)\delta_{s}^{ (0)}(-\bk)\,,
\end{align}
where $\delta_{s}^{ (\ell)} (\bk)= (2\ell+1) \int d^3 x_1 e^{i\bx_1.\bk} \delta_s(\bx_1)\mathcal{L}_\ell(\hat{\bk}.\hat{\bx}_1)$.

\section{Bispectrum Multipole Estimators}\label{sec:bispec}
 
The bispectrum, $B$, is defined via
\begin{align}
\langle  \delta(\bk_1)\delta(\bk_2)\delta(\bk_3)\rangle = (2\pi)^3 \delta_D(\bk_1+\bk_2+\bk_3) B(k_1,k_2,k_3)\,.
\end{align}
For the three point function of the redshift space distorted density fluctuation, the angular dependence $\mu_i\equiv \hat{\bk}_i.\hat{\bn}$ may be regarded as a kinematic dependence, such that one may write 
\begin{align}
\langle \delta(\bk_1;\mu_1)\delta(\bk_2;\mu_2)\delta(\bk_3;\mu_3)\rangle = (2\pi)^3 \delta_D(\bk_1+\bk_2+\bk_3) B(k_1,k_2,k_3;\mu_1,\mu_2,\mu_3)\,.
\end{align}
The most general Legendre decomposition of the angular dependence is then written as
\begin{align}
B^{(\ell_1 \ell_2 \ell_3)}(k_1,k_2,k_3) =\iiint \frac{d\mu_1}{2} \frac{d\mu_2}{2} \frac{d\mu_3}{2}  B(k_1,k_2,k_3;\mu_1,\mu_2,\mu_3)  \left[ \Pi_{i=1}^3(2\ell_i +1)\mathcal{L}_{\ell_i}(\mu_i)\right]\,.
\end{align}
For simplicity, one may instead choose to decompose the angular dependence with respect to only one side, say $k_1$, or equivalently to evaluating 
\begin{align}\label{eq:B_ell_usual}
B^{(\ell)}(k_1,k_2,k_3)\equiv B^{(\ell 0 0)}(k_1,k_2,k_3)\,.
\end{align}

\subsection{Standard Estimator}\label{subsec:StdEst}
The bispectrum
may be estimated at any configuration $( k_1,k_2,k_3)$ by averaging over all configurations within a shell $\pm \Delta k/2$ about each wavenumber, where one restricts to those configurations satisfying the triangle condition. Explicitly the standard estimator is given by
\begin{align}
\hat{B}(k_1,k_2,k_3) = \frac{1}{V_B(k_1,k_2,k_3)}\int_{k_1}  d^3 q_1  \int_{k_2}  d^3 q_2 \int_{k_3} d^3 q_3 \delta_D(\bq_1+\bq_2+\bq_3) \frac{\delta(\bq_1)\delta(\bq_2)\delta(\bq_3)}{V}\,,  
\end{align}
where $\int_{k_1} d^3 q_1$ indicates that we integrate over all wavevectors $\bq_1$ satisfying $q_1\in [k_1\pm \Delta k/2]$, and $V_B(k_1,k_2,k_3)=\int_{k_1}  d^3 q_1  \int_{k_2}  d^3 q_2 \int_{k_3} d^3 q_3 \delta_D(\bq_1+\bq_2+\bq_3)$. By employing the Fourier representation of the Dirac delta function \citep{Sefusatti2005,Fergusson:2010ia,Scoccimarro:2015bla}, $(2\pi)^3 \delta_D(\bk)=\int d^3 \bx \exp(i\bq.\bx)$, one may express the standard estimator in a more computationally efficient form,
\begin{align}
\hat{B}(k_1,k_2,k_3) = \frac{1}{V_B(k_1,k_2,k_3) V}\int \frac{d^3 x}{(2\pi)^3} \left[\Pi_{i=1}^3 \int_{k_i}  d^3 q_i e^{i \bq_i.\bx} \delta(\bq_i)\right]\,,
\end{align}
with the computation for $V_B$ similarly evaluated. Thus the standard estimator is reduced to a three dimensional integral over products of Fast-Fourier Transforms (FFTs). 

\para{Multipoles of the RSD Bispectrum}
The multipole decomposition of the bispectrum is chosen with respect to the largest side, say $k_1$, in the form
\begin{align}
\hat{B}^{(\ell)}(k_1,k_2,k_3) =  \frac{2\ell+1}{V_B(k_1,k_2,k_3)}\left[\Pi_{i=1}^3\int_{k_i}  d^3 q_i\right] \delta_D(\bq_1+\bq_2+\bq_3) \frac{\delta(\bq_1)\delta(\bq_2)\delta(\bq_3)}{V} \mathcal{L}_\ell(\hat{\bk}_1.\hat{{\mathbf{n}}})\,.
\end{align}
One may, of course, reduce the computational burden by (a) the usual replacement of the Dirac delta function, and (b) using the plane parallel approximation to set $\hat{\bf n}\parallel \hat{\bx}_1$, resulting in
\begin{align}\label{eq:Standard_ell_Estim}
\hat{B}^{(\ell)}(k_1,k_2,k_3) = \frac{1}{V_B(k_1,k_2,k_3) V}\int \frac{d^3 x}{(2\pi)^3} \mathcal{D}_{k_1}^{(\ell)}(\bx)  \mathcal{D}_{k_2}^{(0)}(\bx)  \mathcal{D}_{k_3}^{(0)}(\bx)\,, 
\end{align}
where
$
\mathcal{D}_{k}^{(\ell)}(\bx) =  \int_{k}  d^3 q   e^{i \bq.\bx} \delta^{(\ell)}(\bq)\,.
$ This expression may be thought of as the aggregation within subcubes centred at $(k_1,k_2,k_3)$ (of side length $\Delta k$) of the quantity
\begin{align}\label{eq:Bisp_ell_restrict}
\hat{\mathcal{B}}^{(\ell)}(\bk_1,\bk_2,\bk_3) \equiv \frac{ \delta^{(\ell)}(\bk_1) \delta^{(0)}(\bk_2) \delta^{(0)}(\bk_3) }{V}\,.
\end{align}
Despite the efficiency of this algorithm, a potential drawback is the large number of triangles that must be measured unless one coarse-grains to $\Delta k\gg k_f$ (the fundamental frequency $2\pi/V^{1/3}$), with $\mathcal{O}(10^4)$ configurations required otherwise. This coarse graining is required if one is to use mock catalogues to estimate the covariance matrix. The potential drawback is a loss of information. Thus, it is important to check robustness of the estimator using alternative techniques, or, indeed to explore alternative compression techniques.

\subsection{Skew Spectrum}\label{subsec:SkewSpec}
The skew spectrum, developed in \citep{Cooray:2001ps,Szapudi:1997rw,Munshi:1998te,Munshi:2010df}, is motivated by compressing the bispectrum information to a pseudo-power spectrum, such that one may analyse the signal simply by wavenumber. More particularly, one computes the Fourier transform of $\delta(\bx)$ and $(\delta(\bx))^2$ such that
\begin{align}
\langle \delta^2(\bk) \delta(\bk')\rangle = (2\pi)^3 \delta_D(\bk+\bk') S_{\delta^2 , \delta}(k)\,.
\end{align}
It is a trivial task to demonstrate the relation to the underlying bispectrum,
\begin{align}
S_{\delta^2 , \delta}(k) = \int \frac{d^3 q}{(2\pi)^3} B(k , q , |\bk + \bq|)\,.
\end{align}
The skew-spectrum estimator is then computed by evaluating
\begin{align}
\hat{S}_{\delta^2 , \delta}(k) =\int \frac{d^2 \hat{k}}{4\pi} \delta^2 (\bk) \delta(-\bk)\,,
\end{align}
noting that $\delta^2(\bk)$ represents the Fourier transform of $(\delta(\bx))^2$.

It may, however, prove more useful to find a re-weighted form of the skew spectrum, such that, 
\begin{align}
S_{ (w\delta)^2 , w\delta}(k) = \int \frac{d^3 q}{(2\pi)^3} w(k) w(q) w( |\bk + \bq|) B(k , q , |\bk + \bq|)\,.
\end{align}
Similarly, one may extend the definition to take any quadratic form $Q[\delta,\delta]$ in place of $\delta^2$, such that $\langle Q[\delta,\delta](\bk) \delta(\bk')\rangle = (2\pi)^3 \delta_D(\bk+\bk') P_{Q[\delta,\delta] , \delta}(k)$ \citep{Schmittfull:2014tca}. For example, as the tree level SPT bispectrum is given by
\begin{align}
B_{\text{tree}}(k_1,k_2,k_3)=2 F_2(\bk_1,\bk_2)P_L(k_1)P_L(k_2)+ 2\,\text{permutations}\,,
\end{align}
with $P_L$ denoting the linear power spectrum and $F_2(\bk_1,\bk_2)=5/7+\hat{\bk}_1.\hat{\bk}_2/2 (k_1/k_2+k_2/k_1)+2(\hat{\bk}_1.\hat{\bk}_2)^2/7$, then choosing $Q[w\delta, w\delta]=6 \int \frac{d^3 q}{(2\pi)^3} F_2(\bk+\bq,-\bq) \delta(\bk+\bq) \delta(-\bq)$ with weight function $w(k)=1/P_L(k)$, one obtains
\begin{align}
\int \frac{d^3 k}{(2\pi)^3} \hat{S}_{ Q[w\delta, w\delta] , w\delta}(k) = \int \frac{d^3 k}{(2\pi)^3}\frac{d^3 q}{(2\pi)^3}\frac{B_{\text{tree}}(k,q,|\bk +\bq|)}{P_L(k)P_L(q)P_L(|\bk +\bq|)}\delta(-\bk)\delta(-\bq)\delta(\bk +\bq)\,.
\end{align}
As for $\delta^2$, the more general quadratic form can be computed most simply in real space by evaluation of the tidal tensor and derivatives of the density field. The more general form is useful if one wishes to directly compare to a particular model (chosen here as the tree level SPT bispectrum).

\para{Application to RSD Multipoles} The skew-spectrum estimator may be trivially applied to the multipole bispectrum of Eq.~\eqref{eq:Bisp_ell_restrict} to give the estimator
\begin{align}\label{eq:skewspecQ}
S_{Q, \ell}(k)\equiv S_{Q[\delta^{(0)},\delta^{(0)}],\delta^{(\ell)}}(k)
\end{align}
For example, in the standard case that the quadratic form $Q[a,b]$ corresponds  the Fourier convolution $a\star b$, one finds
\begin{align}
S_{\star, \ell}(k)\equiv  S_{ [\delta^{(0)}\star \delta^{(0)}],\delta^{(\ell)}}(k) = \int \frac{d^3 q}{(2\pi)^3} B^{(\ell)} (k , q , |\bk + \bq|)
\end{align}
The estimator is obtained by evaluating the products $\delta^{(0)}$ and $\delta^{(\ell)}$ in real space, and Fourier transforming this quantity to obtain the convolution, and takes the form
\begin{align}
\hat{S}_{\star, \ell}(k)\ =  \int \frac{d^2 \hat{k}}{4\pi} [\delta^{(0)}*\delta^{(0)}] (-\bk)\delta^{(\ell)}(\bk) \,.
\end{align}
As for the standard skew spectrum estimator, it is a simple task to extend this definition to account for arbitrary weighting functions.

\subsection{Line Correlation Function}\label{subsec:LCF}
The line correlation function (LCF), developed in \citep{Obreschkow:2012yb,Wolstenhulme:2014cla}, is a measure of the three point correlation of the phase field  $\epsilon(\bk)=\delta(\bk)/|\delta(\bk)|$ over a scale $r$. Explicitly, one smooths the phase field by convolving with a suitable window function, $W(k|r)$, or equivalently, by evaluating
\begin{align}
\epsilon_r(\bx)=\int \frac{d^3 k}{(2\pi)^3}e^{i\bk.\bx}\epsilon(\bk) W(k|r)\,, 
\end{align} 
where the form of the window function is usually taken to be  a spherical top-hat of the form $\Theta(1-k r/(2\pi))$, where $\Theta$ represents the Heaviside function. Then the LCF for a survey of volume $V$ is given by
\begin{align}\label{eq:LCF_1}
\ell(r) = \frac{V^3}{(2\pi)^9} \left(\frac{r^{3}}{V}\right)^{3/2}  \int \frac{d^2 \hat{r}}{4\pi}\langle \epsilon_r(\bx)\epsilon_r(\bx+\br)\epsilon_r(\bx-\br)\rangle\,.
\end{align}
In terms of the Fourier bispectrum, the line correlation function has leading contribution
\begin{align}\label{eq:LCF_2}
\ell(r)= \left(\frac{r}{4\pi}\right)^{9/2}\iint_{|\bk_1|,|\bk_2|, |\bk_1+\bk_2|\leq 2\pi/r}d^3 k_1 d^3 k_2 \frac{B(k_1,k_2,|\bk_1+\bk_2|)}{\sqrt{P(k_1)P(k_2)P(|\bk_1+\bk_1|)}}  j_0(|\bk_1-\bk_2| r) +\dots
\end{align}
where $\dots$ represents contributions from higher order correlation functions, which vanish in the infinite volume limit, and $j_0$ denotes the zeroth spherical Bessel function.
In order to measure the line correlation function from a given survey or simulation, one evaluates 
\begin{align}\label{eq:LCF_3}
\hat{\ell}(r) =  \left(\frac{r^{3}}{V}\right)^{3/2}\sum_{|\bk_1|,|\bk_2|,\atop |\bk_1+\bk_2|\leq 2\pi/r} \bar{j_0}(|\bk_1-\bk_2|r) \epsilon(\bk_1)\epsilon(\bk_2)\epsilon(-\bk_1-\bk_2)\,,
\end{align}
where $\bar{j_0}(k r)$ is used to denote that the average of $j_0(k r)$ is taken over the volume of a cell of width $k_{\text{f}}$ in Fourier space centred at $\bk$. 

\para{Application to RSD Multipoles}
One may use the same approach directly to evaluate the LCF in the presence of redshift space distortions obtaining
\begin{align}
\hat{\ell}^{(\ell)}(r) =  \left(\frac{r^{3}}{V}\right)^{3/2}\sum_{|\bk_1|,|\bk_2|,\atop |\bk_1+\bk_2|\leq 2\pi/r} \bar{j_0}(|\bk_1-\bk_2|r) \epsilon^{(\ell)}(\bk_1)\epsilon^{(0)}(\bk_2)\epsilon^{(0)}(-\bk_1-\bk_2)\,,
\end{align}
where $\epsilon^{(\ell)} = \delta^{(\ell)}/|\delta^{(\ell)}|$. However, as pointed out in \citep{Eggemeier:2015ifa} the effect of redshift space distortions may be more clearly investigated by decomposing the data into separate components along, and transverse to, the line of sight. This may be achieved by restricting the angle averaging of Eq.~\eqref{eq:LCF_1} to the transverse part of the vector $\br$. This induces the following replacement in Eqs.~\eqref{eq:LCF_2} and~\eqref{eq:LCF_3}:
\begin{align}
j_0(|\bk_1-\bk_2| r) \rightarrow \cos((k_{\parallel,1} - k_{\parallel,2} ) r_\parallel) J_0(|\bk_{\perp,1}-\bk_{\perp,2}| r_\perp)\,,
\end{align}
where $r_\parallel$ and $r_\perp$ denote the parallel and transverse radial separations, $k_{\parallel}=\bk.\hat{\bn}$ and $\bk_\perp\equiv \bk - k_{\parallel}\hat{\bn}$. This form of the line correlation function has been denoted as the `anisotropic line correlation function' (ALCF) \citep{Eggemeier:2015ifa}. This latter approach has shown promising proof-of-concept results using simple Zel'dovich mock fields in determining the Alcock-Paczynski, Kaiser and Fingers-of-God effects. However, a detailed comparison to the results of N-body simulations  has not yet been performed.

\subsection{Integrated Bispectrum}\label{subsec:IB}
The integrated bispectrum was developed in \citep{Chiang:2014oga,Chiang:2015eza} as an efficient method to measure the bispectrum on squeezed configurations. The motivation for this form of the estimator arises from the response function framework; if one divides the survey volume, $V$, into $N_s$ subvolumes (of volume $V_s$), computes the average overdensity, $\bar{\delta}$, within each subcube, and then evaluates the expectation of the product of the local power spectrum, $P(\bk,\br_L)$, with the overdensity over all subvolumes (highlighted with the subscript $N_s$ below), one finds
\begin{align}
\langle P(\bk,\br_L) \bar{\delta}(\br_L)\rangle_{N_s}\rangle =\langle \left(P(\bk)|_{\bar{\delta}=0 }+ \frac{d P(\bk)}{d\bar{\delta}}\Big|_{\bar{\delta}=0}\bar{\delta}(\br_L)+\dots\right)\bar{\delta}(\br_L)\rangle_{N_s} \approx \frac{d \ln P(\bk)}{d\bar{\delta}}\Big|_{\bar{\delta}=0} P(\bk) \sigma_L^2\,,
\end{align}
where $\br_L$ is used to indicate that the relevant quantity is for the subvolume centred at $\br_L$, and where we take the Taylor expansion of the power spectrum in powers of the large scale overdensity, $\bar{\delta}(\br_L)$, with $\sigma_L^2 \equiv  \langle \bar{\delta}^2(\br_L) \rangle_{N_s}$ denoting the variance in mean overdensity over the subvolumes. This quantity is therefore seen to directly probe the `response' of the power spectrum to changes in the large-scale overdensity. Correspondingly,
the integrated bispectrum is defined as
\begin{align}\label{eq:integBisp}
iB(k) = \int \frac{d^2 \hat{k}}{4\pi} \langle P(\bk,\br_L) \bar{\delta}(\br_L)\rangle_{N_s}\rangle\,, 
\end{align}
where one averages over the dependence on orientation which arises if the subvolumes are not isotropic. The relation to the Fourier bispectrum is established simply using the formula for the overdensity,
\begin{align}
\delta(\bk,\br_L)=\int \frac{d^3 q}{(2\pi)^3} \delta(\bk - \mathbf{q})W_L(\mathbf{q}) e^{-i \mathbf{q}.\br_L}\,,
\end{align}
where $W_L(\mathbf{q})= V_s \Pi_{i=1}^3 \text{sinc}(q_i V_s^{1/3}/2)$ is the Fourier transform of the cubic window function defining the subvolume. Substituting $P(\bk,\br_L)=\langle |\delta(\bk,\br_L)|^2\rangle/V_s$ and $\bar{\delta}(\br_L)=\delta(\mathbf{0},\br_L)/V_s$ into Eq.~\eqref{eq:integBisp} one finds
\begin{align}\label{eq:iB_expectation}
iB(k) = \frac{1}{V_s^2} \int \frac{d^2 \hat{k}}{4\pi} \iint \frac{d^3 q_1}{(2\pi)^3} \frac{d^3 q_2}{(2\pi)^3} B(\bk-\mathbf{q}_1,-\bk+\mathbf{q}_1+\mathbf{q}_2,-\mathbf{q}_2) W_L(\mathbf{q}_1) W_L(\mathbf{q}_2) W_L(-\mathbf{q}_1-\mathbf{q}_2)\,.
\end{align}
Measuring the integrated bispectrum is particularly simple, requiring one only to evaluate $\hat{\bar{\delta}}_i$ and $\hat{P}(k)_i$ within each subvolume $i\in [1,N_s]$, and compute the averaged product via
\begin{align}
\widehat{iB}(k) = \frac{1}{N_s} \sum_{i=1}^{N_s} \hat{P}(k)_i \hat{\bar{\delta}}_i\,.
\end{align}

\para{Application to RSD Multipoles} The integrated bispectrum estimator can easily be applied to redshift space distortions, which now corresponds the the response of the power spectrum multipole (c.f. Eq.~\eqref{eq:PowSpecMultipole}) to the monopole overdensity $\bar{\delta}^{(0)}$ for each subvolume, 
\begin{align}
\widehat{iB}^{(\ell)}(k) = \frac{1}{N_s} \sum_{i=1}^{N_s} \hat{P}^{(\ell)}(k)_i \hat{\bar{\delta}}^{(0)}_i\,.
\end{align}
The expectation value is given by Eq.~\eqref{eq:iB_expectation} with $B$ replaced by $B^{(\ell)}$. While the integrated bispectrum is not expected to provide competitive cosmological parameter constraints \citep{Byun:2017fkz}, it is an important probe for squeezed configurations, which may prove particularly useful if one is interested in local primordial non-Gaussianity, for instance.

\subsection{Modal Estimator}\label{subsec:Modal}
The modal estimator involves a decomposition of the bispectrum onto a basis of (symmetric) modes, $Q_n(k_1,k_2,k_3)$, thereby reducing the computation to the evaluation of the coefficients, $\beta_n$, for each mode. The methodology developed first in the context of the CMB in \citep{Fergusson:2009nv,Regan:2010cn} was extended to large scale structure in \citep{Fergusson:2010ia,Regan:2011zq}, with its efficacy demonstrated in \citep{Schmittfull:2012hq} and more particularly, in terms of its potential to provide optimal cosmological constraining power -- in the case of dark matter N-body simulations -- with as few as $\mathcal{O}(10)$ basis modes in \citep{Byun:2017fkz}.

The procedure involves a suitable choice of inner product $\llangle \dots \rrangle$ on the space of all configurations, i.e. those triples $\{ k_1,k_2,k_3 \}$ satisfying $\mathcal{V}=\Big\{ (k_1,k_2,k_3) : 2\, {\text{max}}(k_i) \leq \sum k_i , k_{\text{min}}\leq k_i \leq k_{\text{max}}\Big\}$. Similarly the choice of basis modes is only important in that a suitable choice will improve the rate of convergence, with separability being highly preferred for computational efficiency. With these ingredients in place one wishes to express the bispectrum in the form
\begin{align}\label{eq:bispDecomp}
B(k_1,k_2,k_3) \approx \frac{1}{w(k_1,k_2,k_3)}\sum_n \beta_n Q_n(k_1,k_2,k_3)\,.
\end{align}
for some choice of weighting function, $w$, and basis functions, $Q_n$.
Computation of the basis coefficients may be succinctly expressed in the form
\begin{align}\label{eq:betaCalc}
\beta_n =\sum_m \gamma_{n m}^{-1} \llangle w B | Q_m \rrangle\,, \quad \text{where} \quad \gamma_{nm }= \llangle Q_n | Q_m \rrangle\,.
\end{align}
Estimation of the coefficients $\hat{\beta}_n$ involves the replacement of $B(k_1,k_2,k_3)$ in Eq.~\eqref{eq:betaCalc} by $\delta(\bk_1)\delta(\bk_2)\delta(\bk_3)/V$. In \citep{Fergusson:2010ia} the following inner product was chosen
\begin{align}\label{eq:innerProduct}
\llangle f | g\rrangle = \int \frac{d^3 k_1}{(2\pi)^3}\frac{d^3 k_2}{(2\pi)^3}\frac{d^3 k_3}{(2\pi)^3}(2\pi)^3 \delta_D(\bk_1+\bk_2+\bk_3) \frac{f(\bk_1,\bk_2,\bk_3) g(\bk_1,\bk_2,\bk_3)}{k_1 k_2 k_3}\,,
\end{align}
which, in the case that $f$ and $g$ only depend on the wavenumbers $k_1, k_2, k_3$ reduces to
\begin{align}
\llangle f | g\rrangle = \frac{1}{8 \pi^4}\int_\mathcal{V} dk_1 dk_2 dk_3 f(k_1,k_2,k_3) g(k_1,k_2,k_3)\,.
\end{align}
The latter form is useful for efficient computation of $\gamma_{nm}$.
With this choice of inner product, and defining the weight function as $w(k_1,k_2,k_3)= \\ \sqrt{\frac{k_1 k_2 k_3}{ \hat{P}(k_1)\hat{P}(k_2)\hat{P}(k_3)}}$\,,
the signal to noise in the Gaussian covariance limit may be expressed as
\begin{align}
6 \left( {\mathcal{S} \over \mathcal{N}}\right)^2 =  \llangle w B | w B\rrangle\,. 
\end{align}
While this property is useful to relate covariance of the modes to the non-Gaussian structure induced by non-linear evolution, it is important to highlight that this choice was arbitrary, and any other reasonable choice will work similarly well. The inner product of Eq.~\eqref{eq:innerProduct} may again be written in a computationally efficient manner for separable $f$ and $g$ by utilising the Fourier expansion of the Dirac delta function. Thus the computation reduces to
\begin{align}\label{eq:betaEst}
\hat{\beta}_n = \sum_m \gamma_{n m}^{-1} \llangle  w \hat{\mathcal{B}} | Q_m\rrangle \equiv \sum_{m}\gamma_{n m}^{-1} \frac{1}{V} \int d^3 x \mathcal{M}_{n_1}(\bx) \mathcal{M}_{n_2}(\bx) \mathcal{M}_{n_3}(\bx)\,,  
\end{align}
where $\mathcal{M}_{n_1}(\bx)  = \int \frac{d^3 k}{(2\pi)^3}e^{i\bk.\bx}\frac{q_{n_1}(k)}{\sqrt{k \hat{P}(k)}} \delta(\bk)\,$,
and where we've written $\hat{\mathcal{B}}(\bk_1,\bk_2,\bk_3) \equiv \frac{ \delta(\bk_1) \delta(\bk_2) \delta(\bk_3)}{V}$.
 Here we have utilised the separable expansion of the basis modes -- which as mentioned earlier is especially useful for computational efficiency --,
 \begin{align}\label{eq:Qn}
 Q_n(k_1,k_2,k_3) = \frac{q_{n_1}(k_1)q_{n_2}(k_2)q_{n_3}(k_3) + [\text{5 permutations of the } k_i]}{6}\,.
 \end{align}
 The form of the one dimensional modes $q_n(k)$ is again relatively unimportant, with Fourier modes being perfectly acceptable, though for concreteness, the form of the modes used in \citep{Byun:2017fkz} were polynomials defined to be orthonormal within $\mathcal{V}$ (for details of their construction see \citep{Fergusson:2009nv}).
\para{Application to RSD Multipoles}
The symmetrisation over the basis modes in Eq.~\eqref{eq:Qn} was imposed due to the properties of the bispectrum, $B(k_2,k_3,k_1)=B(k_1,k_2,k_3)$, etc.  
The form of the RSD bispectrum given by Eq.~\eqref{eq:B_ell_usual} is somewhat problematic, as it will not, in general be symmetric.  In Eq.~\eqref{eq:betaEst} we did not need to explicitly symmetrise over the $n_i$ as the arguments of $\mathcal{M}_{n_i}$ were identical. The lack of symmetry in the RSD case originates in the choice of decomposition along one side, $k_1$. One could choose the angle unambiguously -- as advocated in \citep{Hashimoto:2017klo} -- to be the angle between the line of sight and the vector normal to the triangle. Alternatively, one could loosen the symmetry restriction on the basis modes to resolve this issue.
However, with the philosophy that the basis decomposition coefficients could be used in the place of the Fourier bispectrum multipoles for the purpose of parameter constraints rather than reconstruction of the signature on particular configurations, one may instead choose to decompose the quantity
\begin{align}\label{eq:Bisp_ell}
\hat{\mathcal{B}}_{\ell}(\bk_1,\bk_2,\bk_3) \equiv \sum_{\{\ell_i: \ell_1+\ell_2+\ell_3=\ell\}}\frac{ \delta^{(\ell_1)}(\bk_1) \delta^{(\ell_2)}(\bk_2) \delta^{(\ell_3)}(\bk_3) }{V}\,,
\end{align}
which has expectation value 
$
 \sum_{\{\ell_i: \ell_1+\ell_2+\ell_3=\ell\}} B^{(\ell_1 \ell_2 \ell_3)}(k_1,k_2,k_3)\,.
$ We have used a lower subscript to distinguish from the non-symmetric quantity given in Eq.~\eqref{eq:Bisp_ell_restrict} that one decomposes in the standard case.
Due to parity invariance, one need only account for the even multipoles, such that for $\ell=2$ for example the quantity to decompose is given by 
\begin{align}\label{eq:Bisp_ell2}
\hat{\mathcal{B}}_{2}(\bk_1,\bk_2,\bk_3) \equiv \frac{ \delta^{(2)}(\bk_1) \delta^{(0)}(\bk_2) \delta^{(0)}(\bk_3) + 2\,\text{permutations} }{V}\,.
\end{align}
Computation of the modal estimator coefficients $\hat{\beta}_n^{\ell}$ is then trivially inferred as
\begin{align}
&\hat{\beta}_n^{\ell}= \sum_m \gamma_{n m}^{-1} \llangle \hat{\mathcal{B}}_{\ell}| Q_m\rrangle \equiv \sum_{m}\gamma_{n m}^{-1} \frac{1}{V} \int d^3 x \sum_{\{\ell_i: \ell_1+\ell_2+\ell_3=\ell\}}\left(\mathcal{M}^{(\ell_1)}_{\{n_1}(\bx) \mathcal{M}^{(\ell_2)}_{n_2}(\bx) \mathcal{M}^{(\ell_3)}_{n_3\}}(\bx)\right)\,, 
\end{align}
where $ \mathcal{M}_{n_1}^{(\ell)}(\bx)  = \int \frac{d^3 k}{(2\pi)^3}e^{i\bk.\bx}\frac{q_{n_1}(k)}{\sqrt{k \hat{P}(k)}} \delta^{(\ell)}(\bk)\,$, and the shorthand $\{n_1,n_2,n_3\}$ is used to denote that one should take the average over the $6$ permutations\footnote{Clearly for $\ell=2$, Eq.~\eqref{eq:Bisp_ell2} implies that only 3 permutations are necessary, while, for $\ell=0$, just one combination sufficies.} of the $n_i$.

Thus estimation of the RSD bispectrum multipoles is reduced to computation of the basis mode coefficients, $\hat{\beta}_n^{\ell}$. With these coefficients one may reconstruct the decomposed quantity at arbitrary wavenumbers using (c.f. Eq.~\eqref{eq:bispDecomp})
\begin{align}
\hat{\mathcal{B}}_{\ell}(k_1,k_2,k_3) \approx \frac{1}{w(k_1,k_2,k_3)}\sum_n \hat{\beta}_n^{\ell} Q_n(k_1,k_2,k_3)\,.
\end{align}
The data compression represented by these basis coefficients means that computation of the covariance matrix becomes significantly less computationally intensive.

\section{Conclusions}\label{sec:conclu}
As the domain of cosmological research moves ever more into the non-linear regime, the importance of sophisticated prediction and measurement techniques increases. Exploiting the available datasets represented by current and upcoming galaxy surveys requires the measurement of redshift space distortion multipoles of both the power spectrum and bispectrum. In this short paper we significantly increase the number of bispectrum proxies that may be utilised for this purpose. The work of \citep{Scoccimarro:2015bla} demonstrated that existing techniques to efficiently measure the bispectrum for real space simulations may be extended to redshift space distorted galaxy observations. While that work ensured that the complexity involved in measurement of the standard estimator for bispectrum multipoles as with the power spectrum multipoles is reduced to computing FFTs, the vast number of configurations that one must measure -- unless one coarse grains the data considerably -- means that evaluation of the covariance matrix using suites of mock catalogues is unachievable. 

One is, therefore, presented with the option of reducing the number of configurations by coarse graining, at the risk of a loss of information, or searching for alternative proxies. In the former case, it is especially important that robustness checks on the resulting constraints be performed, which again motivates the search for alternative (even if sub-optimal) proxies. Building on \citep{Scoccimarro:2015bla} we have described how real-space estimators for the bispectrum given by the {\it skew spectrum}, the {\it line correlation function}, the {\it integrated bispectrum}, and the {\it modal} (basis) decomposition estimator  may be extended to measure the redshift space distorted bispectrum multipoles. Each of these estimators presents a data compression to a single wavenumber dimension/vector of indices. The basis decomposition method presented by the modal estimator is anticipated from real-space results \citep{Byun:2017fkz} to give competitive constraints to the standard bispectrum proxy. The skew-spectrum may be tuned to search for particular templates (by changing the quadratic form Q, c.f. Eq.~\eqref{eq:skewspecQ}) which can improve its efficiency. By contrast the line correlation function, being somewhat sub-optimal, will primarily be useful as a diagnostic tool. The LCF as a real space, rather than Fourier space, statistic has been demonstrated to be particularly useful as a probe for baryon acoustic oscillations \citep{Eggemeier:2016asq}. In addition a modification of the LCF offers a useful decomposition of the signal into components along, and transverse to, the line of sight. This feature is especially useful for connection to the underlying physical processes. By contrast in \citep{Byun:2017fkz}, it was demonstrated that the integrated bispectrum offers little constraining power for gravitational evolution. However, if one is interested in primordial non-Gaussianity this may become a very useful property, and it is for such purposes that we have also outlined how that proxy may be extended for the case of RSDs. Evaluating the relative efficacy of each proxy is left to future work in which each of these estimators will be applied to galaxy catalogue simulations.

 Confronting upcoming datasets and maximally exploiting the available information requires compression of the signal and a slew of estimation techniques to ensure robustness of the measurements, allied to computational efficiency to extract the available multispectra. The selection of bispectrum multipole proxies presented in this paper offer this promise, as well as computational efficiency. Application of these proxies to realistic surveys will be an important next step in bringing higher order statistics into the realm of standard methods with which to extract useful cosmological signals from galaxy surveys.

\section*{Acknowledgements}
\label{acknow}
The research leading to these results has received funding from the European Research Council under the European Union's Seventh Framework Programme (FP/2007--2013) and ERC Grant Agreement No. 308082.
DR also acknowledges useful discussions with members of the
University of Sussex cosmology group, and especially to David Seery for comments on a draft of this manuscript.

\bibliographystyle{JHEP}
\bibliography{paper}

\end{document}